\documentstyle[12pt]{article}
\begin{document}
\thispagestyle{empty}
\begin{center}
{\LARGE \tt \bf MHD dynamo generation via Riemannian soliton theory}
\end{center}
\vspace{0.5cm}
\begin{center}
{\large \tt \bf L.C. Garcia de Andrade \footnote{Departamento de F\'{\i}sica Teorica, Instituto de F\'{\i}sica,UERJ,Brasil-garcia@dft.if.uerj.br}}
\end{center}
\vspace{2cm}
\begin{abstract}
Heisenberg spin equation equivalence to nonlinear Schr\"{o}dinger equation recently demonstrated by Rogers and Schief , is applied to the investigation of vortex filaments in magnetohydrodynamics (MHD) dynamos. The use of Gauss-Mainard-Codazzi equations allows us to investigate in detail the influence of curvature and torsion of vortex filaments in the MHD dynamos. This application follows closely previous applications to Heisenberg  spin equation to the investigations in magnetohydrostatics given by Schief (Plasma Physics J.-2003). 
\end{abstract}
\noindent
\hspace{2cm} 
\section{Introduction}
 Recently Schief  \cite{1} have shown  that the classical magnetohydrostatic equations of infinitely conducting fluids may be reduced to the integral potential Heisenberg equation constraint to a Jacobian condition as long as the magnetic field is constant along individual magnetic lines. Palumbo's \cite{2} toroidal isodynamic equilibrium has been given as an  example. Earlier Schief has also \cite{3} had provided another very interesting application of how the use of curvature and torsion of lines affects the plasma physical phenomena, by showing that the equilibrium equations of MHD reduce to integral Pohlmeyer-Lund-Regge \cite{4} model subject to a volume preserving constraint if Maxwellian surfaces are assumed to coincide with total pressure constant surfaces. In that paper he provided nested toroidal flux surfaces in magnetohydrostatics. In this paper we provide two new applications of the use of the Heisenberg spin equations to plasma physics. Namely , we apply the Gauss-Mainardi-Coddazzi equations (GMC) \cite{5} to MHD dynamos \cite{6} to compute curvature and torsion effects on vortex filaments of the magnetic field lines. Torsion effects on vortex filaments with and without magnetic fields have been previously investigated by Ricca \cite{7}. More recently Garcia de Andrade \cite{8} have investigated the equilibrium of the magnetic stars more well-known as magnetars. Another interesting example to plasma physics is provided by the Beltrami magnetic flows \cite{9}. These are very important problems in plasma physics and therefore new mathematical methods to address the problem may shed light on their solutions and their applications. This paper seems to be useful to mathematical and plasma physicists. The paper is organised as follows: Section 2 we review the mathematics apparatus of the Serret-Frenet equations and the Heisenberg spin equations. In section 3 we investigate the application of this mathematical framework in explicitely plasma physics problems as the Beltrami magnetic dynamos and the effects of curvature and torsion on vortex filaments.  
\section{Geometrical Heisenberg spin equation}
 In this section we reproduce for the benefit of the reader some of the formulas derived by Rogers and Schief \cite{10} on the Heisenberg spin equation and geometry of curvature and torsion of lines. We begin by defining a Serret-Frenet frame composed of the vectors triad  $X=(\vec{t},\vec{n},\vec{b})$. The extended Serret-Frenet formulas can be written in matrix form are given by
\begin{equation}
\frac{\partial}{{\partial}s}X^{T}= A X^{T}
\label{1}
\end{equation}
where A is given by the array

\vspace{1.0cm}
\displaylines{\pmatrix{0&\kappa&0\cr
	-\kappa&0&\tau\cr
	0&-\tau &0\cr}\cr}\

while the other equations for $\vec{n}$ and $\vec{b}$ direction are given
\begin{equation}
\frac{\partial}{{\partial}n}X^{T}= B X^{T}
\label{2}
\end{equation}
\begin{equation}
\frac{\partial}{{\partial}b}X^{T}= C X^{T}
\label{3}
\end{equation}
where T here represents the transpose of the line matriz X and B and C are the respective skew-symmetric matrices
\vspace{1.0cm}
\displaylines{\pmatrix{0&{\theta}_{ns}&{\Omega}_{b}+{\tau}\cr
	-{\theta}_{ns}&0&-div\vec{b}\cr
	-({\Omega}_{b}+{\tau})&div\vec{b} &0\cr}\cr}
and
\vspace{1.0cm}
\displaylines{\pmatrix{0&-({\Omega}_{n}+{\tau})&{\theta}_{bs}\cr
	({\Omega}_{n}+{\tau})&0&\kappa+div\vec{n}\cr
	-{\theta}_{bs}&-(\kappa+div\vec{n}) &0\cr}\cr}

where ${\theta}_{ns}$ and ${\theta}_{bs}$ are respectively given by
\begin{equation}
{\theta}_{ns}=\vec{n}.\frac{\partial}{{\partial}n}\vec{t}
\label{4}
\end{equation}
and
\begin{equation}
{\theta}_{bs}=\vec{b}.\frac{\partial}{{\partial}b}\vec{t}
\label{5}
\end{equation}
The gradient operator is 
\begin{equation}
{\nabla}= \vec{t}\frac{\partial}{{\partial}s}+\vec{n}\frac{\partial}{{\partial}n}+\vec{b}\frac{\partial}{{\partial}b}
\label{6}
\end{equation}
The other vector analysis formulas read
\begin{equation}
div\vec{t}={\theta}_{ns}+{\theta}_{bs}
\label{7}
\end{equation}
\begin{equation}
div\vec{n}= -{\kappa}+\vec{b}.\frac{\partial}{{\partial}b}\vec{n}
\label{8}
\end{equation}
\begin{equation}
div\vec{b}= -\vec{b}.\frac{\partial}{{\partial}n}\vec{n}
\label{9}
\end{equation}
\begin{equation}
{\nabla}{\times}\vec{t}={\Omega}_{s}\vec{t}+ \kappa\vec{b}
\label{10}
\end{equation}
where
\begin{equation}
{\Omega}_{s}=\vec{b}.\frac{\partial}{{\partial}n}\vec{t}-\vec{n}.\frac{\partial}{{\partial}b}\vec{t}
\label{11}
\end{equation}
which is called abnormality of the ${\vec{t}}$-field. Similarly the results for ${\vec{n}}$ and ${\vec{b}}$ are given  by
\begin{equation}
{\nabla}{\times}\vec{n}= -(div{\vec{b}})\vec{t}+ {\Omega}_{n}\vec{n}
\label{12}
\end{equation}
\begin{equation}
{\Omega}_{n}=\vec{n}.{\nabla}{\times}\vec{n}=-\vec{t}.\frac{\partial}{{\partial}b}\vec{n}-\tau
\label{13}
\end{equation}
and
\begin{equation}
{\nabla}{\times}\vec{b}= ({\kappa}+div{\vec{n}})\vec{t}-{\theta}_{bs}+{\Omega}_{b}\vec{b}
\label{14}
\end{equation}
\begin{equation}
{\Omega}_{b}=\vec{b}.{\nabla}{\times}\vec{b}=-\vec{t}.\frac{\partial}{{\partial}n}\vec{b}-\tau
\label{15}
\end{equation}
\begin{equation}
{\nabla}{\times}\vec{t}={\Omega}_{s}\vec{t}+ \kappa\vec{b}
\label{16}
\end{equation}
where
\begin{equation}
{\Omega}_{s}=\vec{b}.\frac{\partial}{{\partial}n}\vec{t}-\vec{n}.\frac{\partial}{{\partial}b}\vec{t}
\label{17}
\end{equation}
which is called abnormality of the ${\vec{t}}-field$. Similarly the results for ${\vec{n}}$ and ${\vec{b}}$ are given  by
\begin{equation}
{\nabla}{\times}\vec{n}= -(div{\vec{b}})\vec{t}+ {\Omega}_{n}\vec{n}
\label{18}
\end{equation}
\begin{equation}
{\Omega}_{n}=\vec{n}.{\nabla}{\times}\vec{n}=-\vec{t}.\frac{\partial}{{\partial}b}\vec{n}-\tau
\label{19}
\end{equation}
and
\begin{equation}
{\nabla}{\times}\vec{b}= ({\kappa}+div{\vec{n}})\vec{t}-{\theta}_{bs}+{\Omega}_{b}\vec{b}
\label{20}
\end{equation}
\begin{equation}
{\Omega}_{b}=\vec{b}.{\nabla}{\times}\vec{b}=-\vec{t}.\frac{\partial}{{\partial}n}\vec{b}-\tau
\label{21}
\end{equation}
To simplify the magnetic field computations in the next section we shall consider here the particular case of ${\Omega}_{n}=0$ which as has been shown by Rogers and Schief implies the complex lamelar motions and the constancy of magnitude along the streamlines. This geometrical condition implies that the existence of two scalar functions ${\Phi}$ and ${\psi}$ which satisfy the relation
\begin{equation}
\vec{n}={\psi}{\nabla}{\Phi}
\label{22}
\end{equation}
Since tangent planes to the surfaces ${\Phi}=constant$ are generated by the unit tangent $\vec{t}$ and the binormal $\vec{b}$, or  
\begin{equation}
\vec{t}.{\nabla}\Phi= 0
\label{23}
\end{equation}
and
\begin{equation}
\vec{b}.{\nabla}\Phi= 0
\label{24}
\end{equation}
Since $\vec{n}$ is parallel to the normal to surfaces ${\Phi}=const$, the vector lines $\vec{t}$ are geodesics on the surfaces which implies taht the $b-lines$ are parallels on the surface ${\Phi}=const$. The s-lines and b-lines being the parametric curves on the ${\Phi}=const$ surface then a surface metric can be written as 
\begin{equation}
I=d{s^{2}}+g(s,b)db^{2}
\label{25}
\end{equation}
In accordance with the Gauss-Weingarten equations for ${\Phi}=const$ we have the same Serret-Frenet matriz above and  
\begin{equation}
\frac{1}{g^{\frac{1}{2}}}\frac{\partial}{{\partial}b}X^{T}= D X^{T}
\label{26}
\end{equation}
where the matrix D is
\vspace{1.0cm}
\displaylines{\pmatrix{0&-{\tau}&{\theta}_{bs}\cr
	{\tau}&0&\kappa+div\vec{n}\cr
	-{\theta}_{bs}&-(\kappa+div\vec{n}) &0\cr}\cr}
As shown by Rogers and Schief the $\vec{t}-field$ satisfies the Heisenberg spin-type equation
\begin{equation}
\frac{\partial}{{\partial}b}\vec{t}=\frac{\partial}{{\partial}s}(h\vec{t}{\times}\frac{\partial}{{\partial}s}\vec{t})
\label{27}
\end{equation}
where $h=\frac{g^{\frac{1}{2}}}{\kappa}$. The Gauss-Mainardi-Codazzi equations are 
\begin{equation}
{g^{\frac{1}{2}}}\frac{\partial}{{\partial}b}\kappa+\frac{\partial}{{\partial}s}(g\tau)=0
\label{28}
\end{equation}
\begin{equation}
\frac{\partial}{{\partial}b}{\tau}=\frac{\partial}{{\partial}s}[{g^{\frac{1}{2}}}(\kappa+div\vec{n})]+\kappa\frac{\partial}{{\partial}s}{g^{\frac{1}{2}}}
\label{29}
\end{equation}
\begin{equation}
{g^{\frac{1}{2}}}[{\kappa}({\kappa}+div\vec{n})+{\tau}^{2}]=\frac{{\partial}^{2}}{{\partial}s^{2}}{g^{\frac{1}{2}}}
\label{30}
\end{equation}
Besides Rogers and Schief also showed that the Heisenberg spin equation implies the relation
\begin{equation}
\frac{\partial}{{\partial}s}{\kappa}= {\kappa}{\theta}_{bs}
\label{31}
\end{equation}
Most of the expressions revised in this section would be used on the next section in the derivation  of the magnetic field dynamo equations in the Serret-Frenet frame.
\section{MHD dynamos and the Heisenberg equation}
In this section we shall make use of the Salingaros \cite{9} formula for the self-exciting MHD dynamos phenomenologically based, which is expressed as
\begin{equation}
{\nabla}{\times}\vec{B}=k\vec{v}{\times}\vec{B}
\label{32}
\end{equation}
where the magnetic field $\vec{B}=B(s,b,n)\vec{t}$. In principle we have consider the depence of the magnetic field on the Serret-Frenet complete triad lines to later  on simplify this dependence from the field equations of the MHD dynamo. Substitution of this expression for $\vec{B}$ into the MHD dynamo equation (\ref{32}) we obtain  
\begin{equation}
B{\nabla}{\times}\vec{t}+{\nabla}B{\times}\vec{t}=B({\Omega}_{s}\vec{t}+\kappa\vec{b})+\frac{\partial}{{\partial}n}B\vec{b}-\frac{\partial}{{\partial}b}B\vec{n}=kB[v_{n}\vec{b}-v_{b}\vec{n}]
\label{33}
\end{equation}
It is easy to note from this equation that ${\Omega}_{s}=0$ also obtained by Rogers and Schief \cite{12}. This equation of the MHD dynamo should be supplemented by the vanishing monopole condition
\begin{equation}
{\nabla}.\vec{B}= 0
\label{34}
\end{equation}
which imply together the following set of PDE dynamo MHD equations
\begin{equation}
\frac{\partial}{{\partial}n}B=kBv_{n}
\label{35}
\end{equation}
\begin{equation}
\frac{\partial}{{\partial}b}B=-kv_{b}B
\label{36}
\end{equation}
In accordance with the Gauss-Weingarten equations for ${\Phi}=const$ we have the same Serret-Frenet matriz above and  
\begin{equation}
\frac{\partial}{{\partial}s}B = -({\theta}_{bs}+{\theta}_{ns})B
\label{37}
\end{equation}
The Gauss-Mainardi-Codazzi equations and ${\Omega}_{s}=0$ yields 
\begin{equation}
{\nabla}.({\kappa}\vec{b})=0
\label{38}
\end{equation}
Expansion the gradient operator yields
\begin{equation}
\vec{t}\frac{\partial}{{\partial}s}(\kappa(s,b))+\vec{t}\frac{\partial}{{\partial}s}(\kappa(s,b))=0
\label{39}
\end{equation}
since the first term on the LHS of this equation vanishes due to the orthonormality of the Serret-Frenet triad, yields
\begin{equation}
\frac{\partial}{{\partial}b}(\kappa(s,b))=0
\label{40}
\end{equation}
this implies that the curvature depends only on the s-line or $\kappa=\kappa(s)$. Along with equation (\ref{31}) we obtain
\begin{equation}
\frac{\nu}{2}\frac{\partial}{{\partial}s}{\kappa}^{2}=\frac{\partial}{{\partial}b}(\tau)
\label{41}
\end{equation}
\begin{equation}
{\theta}_{bs}=\frac{\partial}{{\partial}b}{\tau}
\label{42}
\end{equation}
where ${\tau}(s,b)$ is the Riemannian torsion which according to Ricca \cite{7} has only something to do with the Cartan torsion tensor when the Serret-Frenet equation are extended to higher dimensions in string theory. Another possibility is that one may find analogies between the Cartan non-Riemannian torsion and the Riemannian one , in Hasimoto soliton like transformation in the Gross-Piraeviskii equation which are the relativistic counterpart of the non-linear Sch\"{o}dinger equation \cite{11}. By considering the ${\Phi}=constant$ surfaces we obtain the following general solution to the MHD dynamo system
\begin{equation}
B(s,b)=e^{[\frac{\partial}{{\partial}b}{\tau}+{\int{v_{b}db}}]}
\label{43}
\end{equation}
Note that the Hasimoto soliton transformation
\begin{equation}
{\psi}={\kappa}(s)e^{[\int{{\tau}ds}]}
\label{44}
\end{equation}
becomes
\begin{equation}
{\psi}={\kappa}(s)e^{[\int{\int{{\theta}_{bs}db}ds}]}
\label{45}
\end{equation}
The mathematical detailed application to dynamos considered here may help to classify the dynamos in general coordinates as has been previously done in part by Salingaros. Although Beltrami spatially periodic fields given by  
\begin{equation}
{\nabla}{\times}\vec{B}=m\vec{B}
\label{46}
\end{equation}
are not self excited fields with $\vec{B}$ generated field by Beltrami flows, for completitude in the next section  we apply the formalism developed by Rogers and Schief to Beltrami fields.
\section{Beltrami fields and flows}
Beltrami magnetic flows are given by 
\begin{equation}
{\nabla}{\times}\vec{v}=k\vec{v}
\label{47}
\end{equation}
Let us consider that the flow velocity is along the s-line direction or $\vec{v}=v_{t}\vec{t}$. One may note that the LHS of equation (\ref{47}) represents the vorticity $\vec{\omega}$ which from Rogers and Schief work \cite{10} yields 
\begin{equation}
\vec{\omega}=v_{t}{\Omega}_{s}\vec{t}+(\frac{\partial}{{\partial}b}v_{t})\vec{n}+(v_{t}{\kappa}-\frac{\partial}{{\partial}n}v_{t})\vec{b}
\label{48}
\end{equation}
Substitution of (\ref{48}) into the Beltrami flow equation yields
\begin{equation}
v_{t}{\Omega}_{s}\vec{t}+\frac{\partial}{{\partial}b}v_{t}\vec{n}+(v_{t}{\kappa}-\frac{\partial}{{\partial}n}v_{t})\vec{b}=m(v_{t}\vec{t})
\label{49}
\end{equation}
Note that the this vectorial equation yields three equations
\begin{equation}
{\Omega}_{s}= m=constant
\label{50}
\end{equation}
\begin{equation}
\frac{\partial}{{\partial}b}v_{t}=0
\label{51}
\end{equation}
\begin{equation}
\frac{\partial}{{\partial}n}v_{t}= {\kappa}v_{t}
\label{52}
\end{equation}
Note that the equation (\ref{48}) is differnt from the equation in the case of MHD dynamo phenomenology proposed by Salingaros, since the abnormality ${\Omega}_{s}$ does not vanish othewise the Beltrami flow would be irrotational or ${\omega}=0$. Applying the divergence operator to the Beltrami flow equation yields
\begin{equation}
{\nabla}.{\nabla}{\times}\vec{v}={\nabla}.\vec{v}
\label{53}
\end{equation}
Since the LHS of the equation (\ref{53}) vanishes the RHS yields
\begin{equation}
{\nabla}.\vec{v}=0
\label{54}
\end{equation}
implies
\begin{equation}
\frac{\partial}{{\partial}s}v_{t}=0
\label{55}
\end{equation}
which along with condition (\ref{51}) shows that $v_{t}$ only depends on the n-line direction. Thus integration of the remaining equation (\ref{52}) yields
\begin{equation}
v_{t}= e^{[-\int{\kappa(n)dn}]}
\label{56}
\end{equation}
Let us now compute the Beltrami magnetic field vortex line. By analogous computations we did in the case of Beltrami flows yields 
\begin{equation}
\frac{\partial}{{\partial}b}B=0
\label{57}
\end{equation}
and
\begin{equation}
\frac{\partial}{{\partial}n}B+{\kappa}B=0
\label{58}
\end{equation}
along with the equation of no magnetic monopole
\begin{equation}
{\nabla}.\vec{B}=0
\label{59}
\end{equation}
yields
\begin{equation}
\frac{\partial}{{\partial}s}B=0
\label{60}
\end{equation}
Therefore the magnetic field in the MHD Beltrami flow is given in the form $\vec{B}(n)$ which from equation (\ref{58}) yields the solution
\begin{equation}
B(n)= e^{[\int{\kappa(n)dn}]}
\label{61}
\end{equation}
which is a very distinct solution of the Salingaros dynamo MHD field.
\section{Conclusions} 
 In conclusion, the effects of curvature and torsion are displayed in MHD dynamos by making use of solitons via teh Heisenberg spin-like equation which is equivalent to Sch\"{o}dinger nonlinear equation where the Hasimoto soliton \cite{12} transformation is performed and using also the mathematical powerful tool of the Serret-Frenet generalised calculus. The comprehension of the geometry and dynamics of vortex filaments in MHD dynamos maybe certainly useful in applications to plasma and solar physics \cite{13}. 
\section*{Acknowledgements}
I would like to thank CNPq (Brazil) for financial support as well a
Universidade do Estado do Rio de Janeiro for financial support. 
\newpage
{\large

\end{document}